\documentclass{llncs}
\usepackage{llncsdoc}
\usepackage{mathptmx,amsmath,epsfig}
\usepackage{amsmath,epsfig}
\usepackage{moresize}
\usepackage{url}
\usepackage[colorlinks=true,
           urlcolor=blue,
           citecolor=blue,
           linkcolor=blue,
           bookmarks=false,
           bookmarksnumbered,
           linktocpage=true
           ]{hyperref}
\usepackage{graphicx}
\usepackage{amsfonts}
\usepackage{amsmath}
\usepackage{subfigure}

\newcommand{\secref}[1]{Section~\ref{#1}}
\newcommand{\figref}[1]{Figure~\ref{#1}}

\newcommand{\mypara}[1]{\vspace{3pt}\noindent\textbf{#1}}

\newcommand{\etal}{\textit{et al.}}

\title{Validation and Inference of Schema-Level Workflow Data-Dependency Annotations} 

\author{Shawn Bowers\inst{1},
Timothy McPhillips\inst{2}, 
Bertram Lud\"{a}scher\inst{2}}
\institute{
Dept.\ of Computer Science, Gonzaga University\\
\and
School of Information Sciences, University of Illinois, Urbana-Champaign
\\[3pt]
{\tt bowers@gonzaga.edu, tmcphillips@absoluteflow.org, ludaesch@illinois.edu}
}

\begin{document}
\maketitle

\begin{abstract}
  An advantage of scientific workflow systems is their ability to
  collect runtime provenance information as an execution trace.
  Traces include the computation steps invoked as part of the workflow
  run along with the corresponding data consumed and produced by each
  workflow step. The information captured by a trace is used to infer
  ``lineage'' relationships among data items, which can help answer
  provenance queries to find workflow inputs that were involved in
  producing specific workflow outputs. Determining lineage
  relationships, however, requires an understanding of the dependency
  patterns that exist between each workflow step's inputs and outputs,
  and this information is often under-specified or generally assumed
  by workflow systems. For instance, most approaches assume all
  outputs depend on all inputs, which can lead to lineage ``false
  positives''. In prior work, we defined annotations for specifying
  detailed dependency relationships between inputs and outputs of
  computation steps. These annotations are used to define
  corresponding rules for inferring fine-grained data dependencies
  from a trace. In this paper, we extend our previous work by
  considering the impact of dependency annotations on workflow
  specifications. In particular, we provide a reasoning framework to
  ensure the set of dependency annotations on a workflow specification
  is consistent. The framework can also infer a complete set of
  annotations given a partially annotated workflow. Finally, we
  describe an implementation of the reasoning framework using
  answer-set programming.
\end{abstract}


\section{Introduction}

Within most scientific workflow systems, a {\em workflow
  specification} (or {\em schema}) is modeled as a graph of nodes
representing computational steps and edges representing the data and
control flow between steps
\cite{davidson08,ludascher2009scientifica}. Each workflow step in a
specification is typically treated as a ``black box'' by the workflow
system. For example, steps are frequently configured to invoke
external programs, execute scripts, or call web services, where the
step exposes only the inputs needed and the corresponding outputs
returned by the underlying calls. Once designed, workflow
specifications serve as executable and potentially reusable (e.g.,
using different input data and parameter settings) scientific
analyses.
Because scientific workflow systems invoke and control the flow of
data between steps during workflow execution, most systems provide
support for recording (or logging) information about a workflow run. A
{\em workflow trace} stores information associated with a run as an
instance of a workflow specification \cite{davidson08,bowers12}. In
particular, traces are modeled as graphs with nodes representing the
invocations of steps and edges representing the data passed between
each step's execution.  Traces are often used to infer the lineage of
workflow data products.  For instance, given a data product output by
a run, many systems use the trace to determine the steps that were
invoked as well as the input and intermediate data products that
contributed to its generation \cite{davidson08,bowers12}.

However, because steps in a workflow specification are black boxes,
workflow systems often ``overestimate'' the lineage relationships
from a workflow trace \cite{bowers12}. For instance, many systems
assume that all data input to a step is used to produce all outputs,
when in fact only a portion of input data may produce any particular
output \cite{cui03,bowers12}. Additionally, most systems consider only
a single, often underspecified notion of dependency between a step's
inputs and outputs, e.g., where data items are said to be ``influenced
by'' or ``contribute to'' other data items \cite{cheney11}. Taken
together, the lineage information inferred from workflow traces may
result in lineage relationships that are not only unclear, but often
misleading or even incorrect.

In 
\cite{bowers12}, we developed a set of declarative rules  
for specifying dependency patterns of individual computation
steps. The inputs and outputs of a step are annotated with rules,
which are then used to infer the specific input data used to produce
an output for each invocation of a step within a trace. However, to be
effective, this approach requires a complete set of annotations for
every step within a workflow specification.

\mypara{Contributions.} 
We describe extensions to our
prior work that supports partially annotated workflow specifications
and employs reasoning techniques to validate and help infer a complete
set of annotations. We consider different use cases
related to annotating a workflow specification and provide a set of
dependency types that can be used to help clarify the lineage
relationships present within a workflow trace. 
Finally, we describe a prototype implementation (using answer-set
programming) of our approach that we plan to add to the YesWorkflow
system \cite{mcphillips15} as future work.

\mypara{Organization.} 
In \secref{sec:overview} we describe an abstract model of workflow
specifications, give an overview of the dependency types we consider
for annotations, and discuss use cases related to our framework.  In
\secref{sec:model} we describe the constraints associated with the
dependency types as well as the corresponding inferences for reasoning
over partially annotated workflow specifications. In
\secref{sec:prototype} we present a prototype implementation of the
reasoning approaches described in \secref{sec:model}. Finally, in
Sections \ref{sec:related} and \ref{sec:conclusion} we describe related and
future work, respectively.

\section{Workflow Dependency Annotations}
\label{sec:overview}

This section describes an abstract model for workflow specifications
used in the rest of the paper, an overview of the types of
dependencies we consider for workflow annotations, and three example
use cases related to annotation inference.

\begin{figure}[!t]
  \vspace{-20pt}
\centerline{\includegraphics[scale=.75]{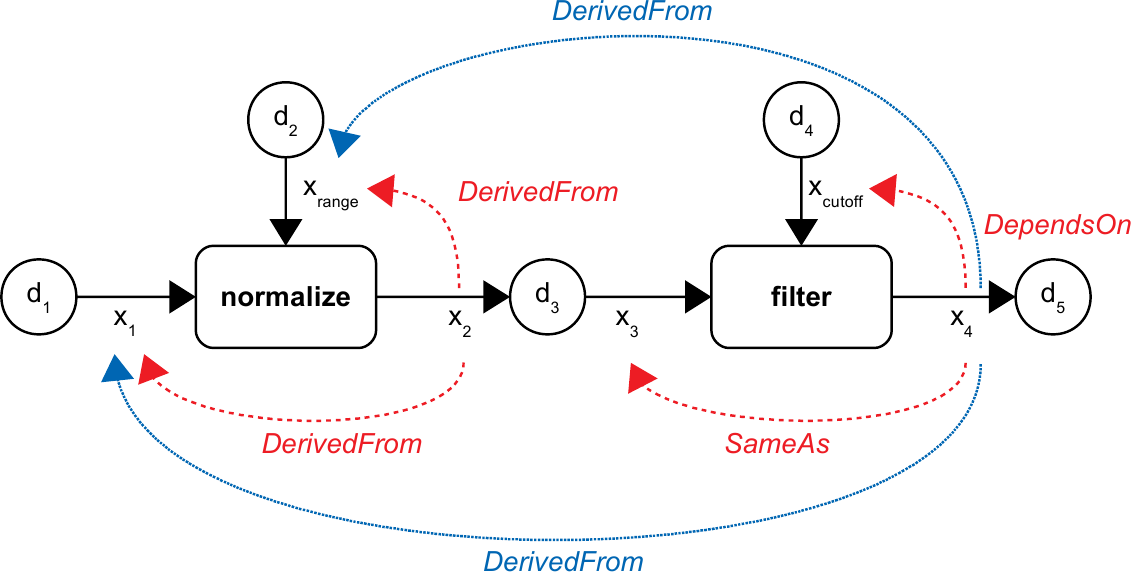}}
  \vspace{-8pt}
  \caption{Example workflow with program blocks \texttt{normalize} and
    \texttt{filter}, data blocks $d_1,\dots,d_5$, and dataflow edges
    (solid, black) between nodes;  user-declared dependency
    annotations (dashed, red edges); and inferred dependencies
    (dotted, blue edges), based on the given user annotations.}
  \label{fig:example1}
\end{figure}

\mypara{Workflow Specifications.} 
We consider an abstract workflow model that conforms to YesWorkflow
 and similar dataflow-oriented scientific workflow
models \cite{davidson08,bowers12}.  A workflow $W = (P, D, E)$
consists of a set of {\em program blocks} $P$ (workflow steps, i.e.,
computations), {\em data blocks} $D$ (representing data items or data
containers), and {\em input} and {\em output} edges
$E \subseteq P \times L \times D \times \{{\tt in},{\tt out}\}$ where
$L$ is a set of labels that uniquely identify edges within $W$.
We use relations ${\tt in}(p_i,x_i,d_i)$ and ${\tt out}(p_j,x_j,d_j)$
to denote input and output edges, respectively, for $p_i,p_j \in P$,
$x_i,x_j \in L$, and $d_i,d_j \in D$. \figref{fig:example1} shows an
example workflow consisting of two program blocks ({\tt normalize} and
{\tt filter}), five data blocks ($d_1, \dots, d_5$), four input edges
($x_1$, $x_3$, $x_{\mathtt{range}}$, and $x_{\mathtt{cutoff}}$), and
two output edges ($x_2$ and $x_4$). Also shown in
\figref{fig:example1} are a set of initial dependency annotations
(red, dashed) together with the corresponding inferred annotations
(blue, dotted). The {\tt normalize} block takes input data items $d_1$
and scales them to fit within the given range (consisting of a minimum
and a maximum value). The output of {\tt normalize} is then passed to
{\tt filter}, which outputs the data item if it is smaller than a
given cutoff value $d_4$.  In general, an input edge
${\tt in}(p_1,x_1,d_1)$ states that data items are input to the
program block $p_1$ and an output edge ${\tt out}(p_1,x_2,d_2)$ states
that data items are output by $p_1$.
Data blocks allow for data items to be passed as input to multiple
program blocks (e.g., to create workflow branches as in $d_2$ in
\figref{fig:example4}). In contrast, data blocks typically receive
only from a single writer, to avoid conflicts (e.g., $d_3 \neq d_4$ in
\figref{fig:example4}).\footnote{If data blocks denote containers
  (e.g., file folders or queues) multiple writers may be allowable.}

\sloppypar \mypara{Dependency Annotations.} The set of dependency
annotations $A \subseteq L \times L \times T$ for a workflow
specification $W$ associates different dependency types $t \in T$ to
input and output edges of $W$. Dependency annotations are represented
by a relation ${\tt dep\_rule}(x_1,x_2,t)$ for input edges
$x_1 \in L$, output edges $x_2 \in L$, and dependency types $t \in T$.
The dashed, red arrows in \figref{fig:example1} represent four
explicit, user-supplied annotations:
${\tt dep\_rule}({x}_1,{x}_2, \mathtt{DerivedFrom})$,
${\tt dep\_rule}(x_\mathtt{range},{x}_2, \mathtt{DerivedFrom})$,
${\tt dep\_rule}({x}_3,{x}_4, \mathtt{SameAs})$, and
${\tt dep\_rule}(x_\mathtt{cutoff},{x}_4, \mathtt{DependsOn})$. In
the example, we say that the output of {\tt normalize} is ``{\em
  derived from}'' the input ${d}_1$ and the {\tt range} ${d}_2$, and
the output of {\tt filter} ``{\em depends on}'' the {\tt cutoff}
${d}_4$ and is the ``{\em same (data item) as}'' the input ${d}_2$. We
note that annotations can be expressed over a single program block
(e.g., the explicit annotations in \figref{fig:example1}) or can span
multiple program blocks (e.g., the inferred annotations in
\figref{fig:example1}).

\mypara{Dependency Types.}  We consider a set of pairwise disjoint
dependency types for specifying annotations. The {\em FlowsFrom} type
simply represents the cases where an input data item was received and
an output item was produced by a program-block invocation, but the
output value is not determined by or computed from the input. A {\em
  FlowsFrom} annotation typically denotes that the input is simply a
``trigger'' to tell the program block to be invoked. The {\em
  DependsOn} type represents cases where a control dependence exists
between the corresponding inputs and outputs (explained in more detail
in \secref{sec:model}). The {\em DerivedFrom} type represents cases
where outputs are computed from inputs (again, described further in
\secref{sec:model}). The {\em ValueOf} type represents the cases where
an output produces a new data item (with a new object identifier)
containing a copy of the input data item's value.  Finally, the {\em
  SameAs} type represents the cases where the input data item was
passed through to the output (i.e., the output is the same exact data
item as the input data item).

\mypara{Use Case 1: Inferring Dependency Annotations.}  Given a
workflow specification that is partially annotated, we consider the
case of inferring new annotations from a given set of user-supplied
annotations.  \figref{fig:example1} gives a simple example where each
program block is annotated (dashed, red arrows) and the corresponding
annotations that are implied by the given annotations are also shown
(blue arrows). In this example, each individual workflow step is
annotated by a user, and the goal is to infer the annotations that
span multiple steps.  In general, understanding the dependency
relationships that span workflow steps as a result of the composition
of program blocks is useful for verifying the intent and/or
construction of the workflow (e.g., to ensure that certain workflow
outputs are actually derived from certain workflow inputs). Having a
complete set of annotations is also useful when answering queries at
the trace level, e.g., to determine the inputs that specific outputs
were derived from (as opposed to the inputs that were simply copied
from the input or were used for basic control flow).

\begin{figure}[!t]
  \vspace{-20pt}
  \centerline{\includegraphics[scale=.75]{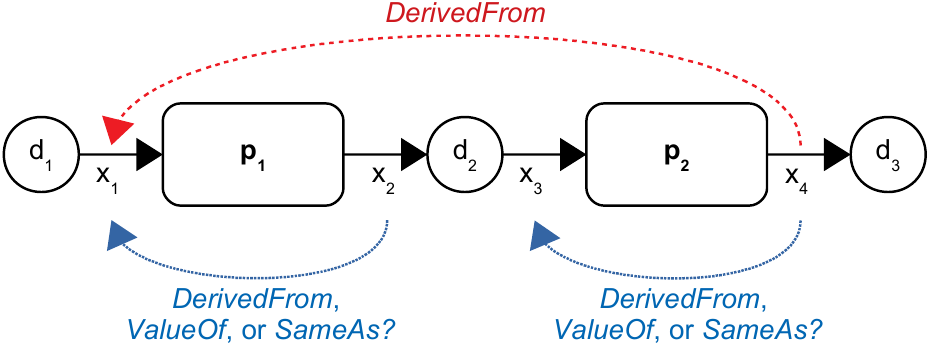}}
 \vspace{-8pt}
 \caption{Example workflow with initial user annotation (dashed, red)
   from the output $x_4$ to the input $x_1$.  Which of the undeclared
   dependency options (dotted, blue) are correct?}
  \label{fig:example2}
\end{figure}

\mypara{Use Case 2: Constraining Dependency Annotations.} In this
case, higher-level annotations that span multiple program blocks
(e.g., between workflow inputs and outputs) are used to help guide
annotation choices for the rest of the workflow specification. As a
simple example, we may know that the output is (or should be) derived
from the input as shown in \figref{fig:example2} by the dashed red
annotation.  Specifying this annotation first limits the choices for
the lower-level annotations (in this case of program blocks). The
corresponding choices are shown by the dotted blue annotations in
\figref{fig:example2}. In this case, different combinations of
annotations over the two program blocks are compatible (consistent)
with the initial (dashed red) annotation of \figref{fig:example2}.

\begin{figure}[!t]
  \vspace{-20pt}
\centerline{\includegraphics[scale=.75]{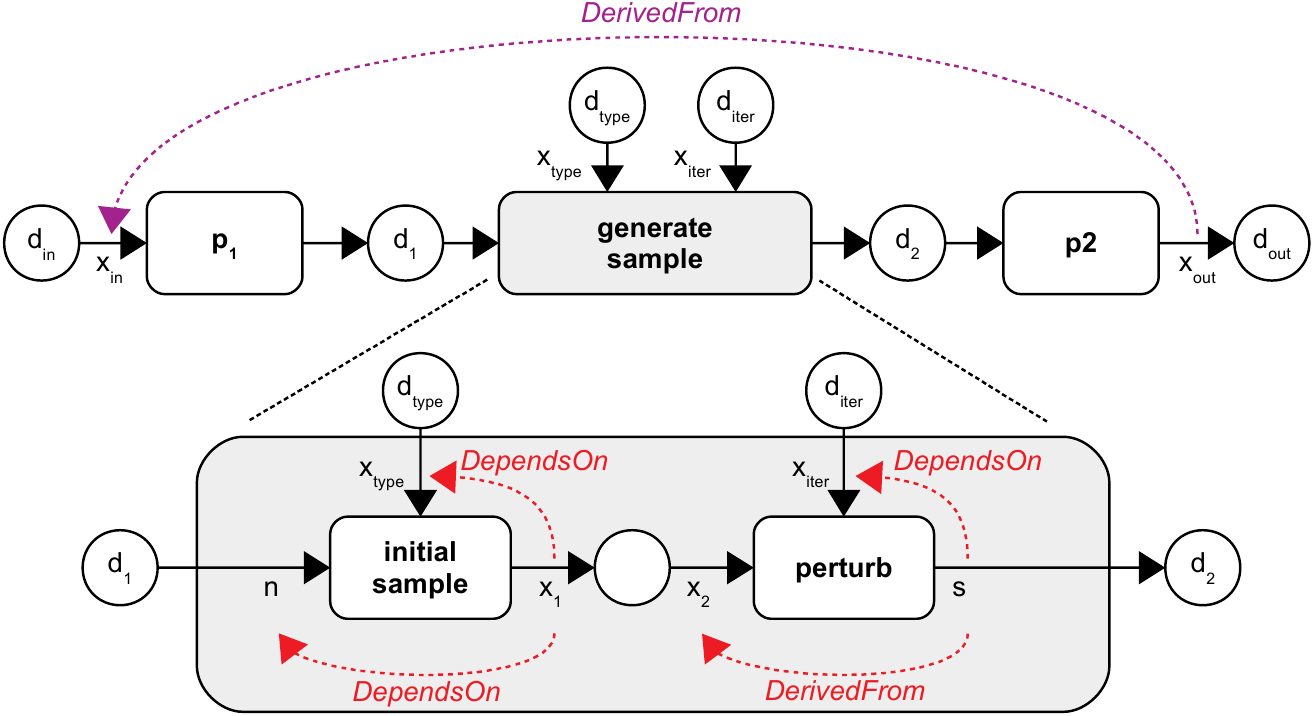}}
  \vspace{-10pt}
  \caption{Workflow specification consisting of an annotated
    subworkflow (dashed red, bottom) and an inconsistent higher-level
    annotation assertion (dashed purple, top) that spans workflow
    steps.}
  \label{fig:example3}
\end{figure}

\mypara{Use Case 3: Validating Dependency Relationships.} Finally, we
consider the case where there is a mix of (potentially partial)
higher-level (i.e., indirect) and lower-level (i.e., direct)
annotations of a workflow specification that a workflow designer wants
to ensure are compatible (consistent). \figref{fig:example3} is one
such example where the workflow specification consists of a
subworkflow (named {\tt generate\_sample} as shown on the bottom of the
figure). Each subworkflow step is annotated (in red) and the
containing workflow (shown on the top of the figure) has a
higher-level annotation asserting that the output should be derived
from the input. However, the given annotations are incompatible
(i.e., inconsistent) since the composition of the two subworkflow
steps introduce an implied {\em DependsOn} relationship between the
input and output of {\tt generate sample}. Thus, based on the workflow
specification, ${d}_{\tt in}$ and ${d}_{\tt out}$ cannot participate
in a \emph{DerivedFrom} relationship (as shown at the top in purple).

\vspace{6pt}
The reasoning framework we describe in the rest of this paper is
designed to handle each of these three cases. In particular, we assume
that a workflow specification is either fully or partially annotated,
from which the reasoning framework (i) ensures consistency of the
given annotations (e.g., as in \figref{fig:example3}); (ii) infers all
specific implied annotations (e.g., as in \figref{fig:example1}); and
(iii) provides the allowable annotation options when there are
multiple possible implied annotations (e.g., as in
\figref{fig:example2}).


\section{Reasoning over Dependency Types}
\label{sec:model}

This section describes our reasoning framework for dependency type
validation and inference. We first give a more detailed description of
the annotation types and then describe the annotation composition
rules and constraints used within our framework.

\subsection{Dependency Types}

In the following, we assume a simple program block $p$ with input edge
${\tt in}(p,x_1,d_1)$ and output edge ${\tt out}(p,x_2,d_2)$ as shown
below.
\begin{center}
  \includegraphics[scale=.75]{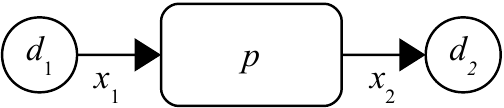}
\end{center}
Let $D_1$ be the set of allowable values (the \emph{domain}) of $p$
with respect to the input edge $x_1$ and $D_2$ be the set of possible
output values (the \emph{range}) with respect to the output edge
$x_2$.  We write $p : D_1 \to D_2$ to denote the {\em signature} of
$p$ with respect to $x_1$ and $x_2$. We assume data items are passed
to and from program blocks as objects $o$ with unique identifiers
$id(o)$ and corresponding values $val(o)$. For a domain $D$ and a
sequence of data items $\bar{o}$, we write $val(\bar{o}) \subseteq D$
if for every data item $o_i \in \bar{o}$, $val(o_i) \in D$. Given the
program block signature $p:D_1\to D_2$, an {\em invocation}
$p(\bar{o}_1) = \bar{o}_2$ states that $p$ read a sequence of data
items $\bar{o}_1$ on $x_1$ such that $val(\bar{o}_1) \subseteq D_1$,
and wrote a (possibly empty) sequence of data items $\bar{o}_2$ on
$x_2$ such that $val(\bar{o}_2) \subseteq D_2$.\footnote{The
  use of sequences of data items allows for more complex program
  blocks such as filters and aggregators as well as workflow
  computation models supporting implicit iteration
  \cite{alper18,bowers12}.} Program blocks are not required to be
deterministic, and so different invocations over the same input may
produce different output. The {\em image} $p[\bar{o}_1]$ of
$\bar{o}_1$ under $p$ is the set of all possible output sequences
produced by invocations of $p$ receiving $\bar{o}_1$. Note that if $p$
has multiple input edges, the same notion of image still applies since
we are interested in the relationship between a single input and
output edge (although additional constraints are imposed in some cases
as described below).

Following the traditional convention used in programming language
implementation \cite{ferrante87,cheney11}, we use the ideas of
``control'' and ``data'' dependence between statements when defining
the dependency types below. For example, consider the following
statements (adapted from \cite{ferrante87}).
\begin{footnotesize}
\begin{verbatim}
   S1: C = A * B
   S2: E = C * D + 1
   S3: if (E > 0) then
   S4:    H = F + G
\end{verbatim}
\end{footnotesize}
Statement {\tt S2} is said to have a {\em data dependence} on {\tt S1}
since the value of {\tt E} depends on the value of {\tt C}. A data
dependence is also referred to as a ``{\em read-after-write}''
dependence since {\tt C} is read as part of {\tt S2} to compute a
value to write to {\tt E}. Note that data dependence relationships can
either be direct or indirect. For instance, in the example above, {\tt
  E} directly depends on {\tt C} (via \texttt{S2}) but indirectly
depends on {\tt A} (via \texttt{S1} and \texttt{S2}).  Below, we write
${\tt raw\_dep}(p,x_1,x_2)$ to denote that within a program block $p$,
output edge $x_2$ has either a direct or indirect read-after-write
dependence on input edge $x_1$. Similarly, statement {\tt S4} is said
to have a {\em control dependence} on statement {\tt S3} since the
execution of statement {\tt S4} (and hence, the value of {\tt H})
depends on the execution of {\tt S3} (specifically, the value of {\tt
  E}).  However, note that {\tt H}'s value is not computed from {\tt
  E}'s value (which would imply a data dependence). A control
dependence can also be either direct or indirect. We assume that if an
$x_2$ is indirectly control dependent on $x_1$ then either: (i) $x_2$
is control dependent on another variable that is either directly or
indirectly control or data dependent on $x_1$; or (ii) $x_2$ is data
dependent on a variable that is either directly or indirectly control
dependent on $x_1$.
Below, we write ${\tt ctl\_dep}(p,x_1,x_2)$ to denote that $x_2$ has
either a direct or indirect control dependence on $x_1$.  We define
the dependency types below in terms of the constraints they impose
between possible inputs and outputs of program-block invocations as
well as their corresponding control and data dependences.

\mypara{FlowsFrom}.  A {\em FlowsFrom} annotation implies that $x_2$
does not have a control or data dependence on $x_1$, which is
expressed by the constraint:
$$\neg\, {\tt ctl\_dep}(p,x_1,x_2) \;\wedge\; \neg\,
{\tt raw\_dep}(p,x_1,x_2).$$ {\em FlowsFrom} simply suggests that the
input was present when $p$ was executed, e.g., the input was used as a
``trigger'' to invoke a program block $p$.

\mypara{DependsOn}. A {\em DependsOn} annotation implies that $x_2$
has a control dependence, but not a data dependence on $x_1$, which
is expressed by the constraint:
$${\tt ctl\_dep}(p,x_1,x_2) \;\wedge\; \neg\; {\tt raw\_dep}(p,x_1,x_2).$$

\mypara{DerivedFrom}. A {\em DerivedFrom} annotation implies that
$x_2$ has a data dependence on $x_1$, but that not all outputs have
the same value(s) as their corresponding inputs (which would suggest a
{\em ValueOf} or {\em SameAs} relationship):
$${\tt raw\_dep}(p,x_1,x_2) \;\wedge\; \, (\exists \bar{o}_2  \in p[\bar{o}_1] \,{:}~ val(\bar{o}_2) \not\subseteq val(\bar{o}_1)).$$ 
As explained further below, we consider {\em DerivedFrom} to be a
``stronger'' dependency relationship than {\em DependsOn}. Thus, while
it is possible for $x_2$ to have both a control and data dependence on
$x_1$, it would be represented as {\em DerivedFrom} within our
framework.

\mypara{ValueOf}. A {\em ValueOf} annotation implies that the values
of data items received on $x_1$ are output on $x_2$ (e.g., by copying
inputs to new outputs). Unlike with {\em SameAs}, {\em ValueOf}
assumes new data items are created as a result, and so the identifiers
for the input and output data items differ:
$$(\forall \bar{o}_2  \in p[\bar{o}_1] \,{:}~  val(\bar{o}_2) \subseteq val(\bar{o}_1)) 
~\wedge~ (\exists \bar{o}_2 \in p[\bar{o}_1] \,{:}~ id(\bar{o}_2)
\not\subseteq id(\bar{o}_1)).$$ We use $id(\bar{o})$ to denote the set
of identifiers of the sequence of data items $\bar{o}$. Note that {\em
  ValueOf} implies a data dependence from $x_2$ to $x_1$ since data
items must be read from input $x_1$ and then written into data items
that are output to $x_2$.

\mypara{SameAs.}  A {\em SameAs} annotation differs from {\em ValueOf}
by requiring all outputs to be the same as data items from the inputs:
$$\forall \bar{o}_2 \in p[{\bar o}_1] \,{:}~  o \in \bar{o}_2 \to o \in \bar{o}_1.$$
Here, $o \in \bar{o}$ holds if the object $o$ is a member of the
sequence $\bar{o}$. A {\em SameAs} relationship also implies a data
dependence from $x_2$ to $x_1$ since the input data items must be read
from $x_1$ and then written to $x_2$.

\subsection{Composing Dependency Annotations}

Annotation inference within a workflow specification is largely based
on understanding how annotations ``propagate'' under compositions (or
``sequences'') of workflow steps.  Here we assume two connected
program blocks $p_1 : D_1 \to D_2$ and $p_2 : D_2 \to D_3$:
\begin{center}
  \includegraphics[scale=.75]{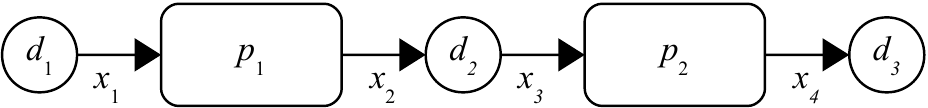}
\end{center}
When $p_1$ and $p_2$ are connected by a data block as above,
we write $p_1 \circ p_2$ to denote the connection. We also define the
ordering $\prec$ to represent the intuitive ``dependency strength'' of
annotation types. In particular, if $t_i \prec t_j$ then we say $t_i$
is a ``weaker'' dependency type than $t_j$ (or similarly, that $t_j$
is a ``stronger'' dependency type than $t_i$).
The dependency types are ordered according to dependency strength as
follows.
\begin{center}
  {\em FlowsFrom} $\prec$ {\em DependsOn} $\prec$ {\em
    DerivedFrom} $\prec$ {\em ValueOf} $\prec$ {\em SameAs}
\end{center}
For instance, a {\em DependsOn} relationship suggests a ``weaker''
dependency than a {\em DerivedFrom} relationship. The definitions of
the annotation types with the ordering above imply the following
annotation composition rules for a sequence of program blocks
$p_1 \circ p_2$, with ${\tt in}(p_1,x_1,d_1)$,
${\tt out}(p_1,x_2,d_2)$, ${\tt in}(p_2,x_3,d_2)$, and
${\tt out}(p_2,x_4,d_3)$ as defined above, and $\preceq$ denoting
weaker or of equal strength (and where all variables are assumed below
to be universally quantified).
\begin{equation*}
\begin{array}{c}
  {\tt dep\_rule}(x_1,x_2,t_i) \wedge {\tt dep\_rule}(x_3,x_4,t_j)
  \wedge t_i \preceq t_j \leftrightarrow {\tt dep\_rule}(x_1,x_4,t_i) \\[3pt]
  {\tt dep\_rule}(x_1,x_2,t_j) \wedge {\tt dep\_rule}(x_3,x_4,t_i)
  \wedge t_i \preceq t_j \leftrightarrow {\tt dep\_rule}(x_1,x_4,t_i) 
\end{array}
\end{equation*}
These rules can also be applied to indirect annotations (spanning
multiple blocks) as well, which is further described in
\secref{sec:prototype}. As an example of propagation, in
\figref{fig:example1}, {\tt normalize} has a {\em DerivedFrom}
annotation and {\tt filter} has a {\em SameAs} annotation. Since {\em
  DerivedFrom} is ``weaker'' than {\em SameAs}, the composite
annotation is {\em DerivedFrom}.
Similarly, in \figref{fig:example2} the composite annotation is {\em
  DerivedFrom}, which implies that $p_1$ and $p_2$ have either {\em
  DerivedFrom} annotations or ``stronger'' types (i.e., {\em ValueOf}
or {\em SameAs}), since {\em DerivedFrom} must be the ``weaker''
annotation.  Additionally (and not shown in \figref{fig:example2}),
note that at least one of $p_1$ or $p_2$ must have a {\em DerivedFrom}
annotation to satisfy the composition rules above. The example in
\figref{fig:example3}, while slightly more complex, follows the same
idea in that along the path from ${\tt x}_{\tt out}$ to
${\tt x}_{\tt in}$, the {\tt generate\_sample} subworkflow implies a
{\em DependsOn} annotation, and since {\em DependsOn} is strictly
weaker than {\em DerivedFrom}, the higher-level {\em DerivedFrom}
annotation violates (is inconsistent with) the composition rules.

According to the composition rules, weaker annotations propagate
through program-block compositions, which is due to the nature of the
dependencies established by the weaker annotation. For instance, if
$x_2$ {\em FlowsFrom} $x_1$, then $d_2$ (via $x_2$) does not have a
control or data dependence on $d_1$ (via $x_1$).  Thus, since the
value of $d_1$ does not participate in the computation of $d_2$, $d_1$
also does not participate in the computation of the values that have a
control or data dependence on $d_2$. A similar situation exists when
$p_2$ has a {\em FlowsFrom} annotation. Determining indirect control
dependences (i.e., when looking at sequences of statements involved in
control and data dependences) was described in the beginning of this
section, and follows from the idea that control dependence can be
indirectly established through other control and/or data
dependences. The same ideas apply to copying the values of
data items. If $d_2$ is a (value) copy of $d_1$ with potentially
different data item identifiers as $d_1$ (i.e., $x_2$ has a {\em
  ValueOf} relationship with $x_1$), but $d_2$ is passed through to
$d_3$ (i.e., $x_4$ has a {\em SameAs} relationship with $x_3$), then
$d_3$ will also have the same value but a different identifier as
$d_1$ (since $d_2$ and $d_3$ are the same data item).  The same
situation occurs when the two annotations are flipped, i.e., $p_1$ has
a {\em SameAs} relationship and $p_2$ has a {\em ValueOf}
relationship.  Finally, when $p_1$ and $p_2$ have the same exact
annotation, the same annotation is also propagated, which follows from
similar arguments as those above.

\subsection{Additional Annotation Constraints}

\begin{figure}[!t]
  \begin{center}
    \includegraphics[scale=.75]{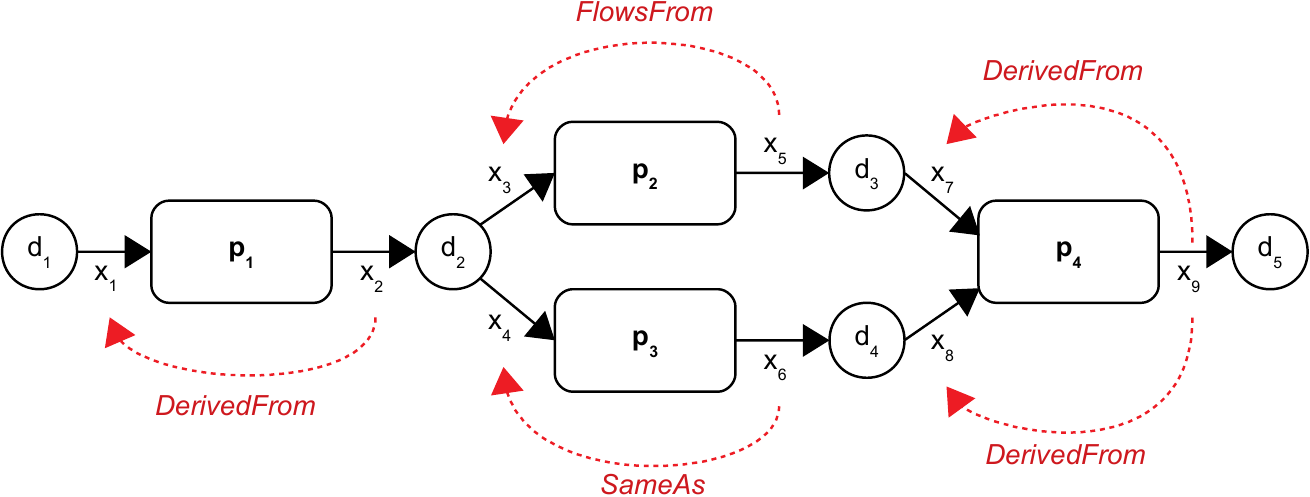}
  \end{center}
  \vspace{-16pt}
  \caption{Example workflow specification with multiple paths between
    the input and output.}
  \label{fig:example4}
\end{figure}

We also consider an additional ``global'' constraint on the dependency
annotations of a workflow specification related to inferring
annotations when there are two or more paths of program-block
compositions within a workflow specification. Consider the example
annotated workflow specification of \figref{fig:example4}, which shows
two paths (i.e., sequences of program block compositions) between
$x_1$ and $x_9$. While the top path (through $p_2$) implies a {\em
  FlowsFrom} relationship from $x_9$ to $x_1$ (since {\em FlowsFrom}
is the weakest type along the path), the bottom path implies a
stronger {\em DerivedFrom} relationship from $x_9$ to $x_1$. Since we
allow at most one dependency type between an input and an output,
we use the annotation inferred from the path with the strongest
type.

\section{Prototype Implementation}
\label{sec:prototype}

This section describes a prototype implementation of our annotation
reasoning framework using the Potassco\footnote{See:
  \url{https://potassco.org/}} suite of answer-set programming (ASP)
tools. Potassco implements ASP using a syntax similar to Datalog with
additional support for nonmonotonic reasoning based on the answer set
semantics \cite{gelfond14}. Potassco programs are often written using
a generate-and-test algorithmic approach where the result of a program
is a set of minimal models, or ``{\em answer sets}'', that satisfy the
rules and constraints defined within the program. Our implementation
follows this same approach by:

(i) ``guessing'' dependency annotations for each input-output pair in
a workflow specification without a corresponding user-supplied
annotation (the generate step);

(ii) ensuring that each of the input-output pair annotations satisfy
the program-block annotation compositions described in the previous
section (the test step); and

(iii) ensuring that annotations satisfy the additional constraints
described in the previous section, i.e., ensuring the ``strongest''
indirect annotations are used between inputs and outputs with multiple
paths of program blocks between them (the test step).

In the generate-and-test approach, conceptually all possible models
are created---which in our case means that all possible combinations
of input-output pair combinations along a dataflow path are
considered---and only those models (answer sets) that satisfy the
given constraints are returned.
Our prototype implementation uses the answer sets for a workflow
specification and then (i) outputs all annotations that are contained
in each answer set (i.e., the annotations that are ``entailed'' by
the program); and then (ii) outputs the annotation choices (i.e., the
union of annotations across answer sets) for the annotations that are
not entailed (e.g., as is the case with the blue annotations in
\figref{fig:example2}).

Our prototype uses a ``choice rule'' to generate annotations for
input-output pairs not already annotated as part of the workflow specification:
\begin{footnotesize}
\begin{verbatim}
  {dep_rule(I,O,R) : dep_type(R)} = 1 :- up_stream(I,O).
\end{verbatim}
\end{footnotesize}
Where {\tt up\_stream(I,O)} finds all potential input-output
annotation pairs:
\begin{footnotesize}
\begin{verbatim}
  up_stream(I,O) :- in(I,P,_), out(O,P,_).
  up_stream(I,O) :- in(I,P1,_), out(O1,P1,D1), in(I2,P2,D1), up_stream(I2,O).
\end{verbatim}
\end{footnotesize}
The following constraint ensures that all annotations satisfy the
composition rules:
\begin{footnotesize}
\begin{verbatim}
  :- dep_rule(I,O,R), not valid_dep_path(I,O,R).
\end{verbatim}
\end{footnotesize}
In ASP the head of the (constraint) rule above is assumed to be
false. Thus, if the body is satisfied the constraint fails. To satisfy
the constraint, the body must not be true. So, in the constraint
above, either there does not exist a dependency between the input {\tt
  I} and output {\tt O}, or the dependency forms a valid dependency
path. The relation {\tt valid\_dep\_path(I,O,R)} is true if there is a
valid annotation with type {\tt R} between the input {\tt I} and
output {\tt O} as defined below.
\begin{footnotesize}
\begin{verbatim}
  valid_dep_path(I,O,R) :- in(I,P,_), out(O,P,_), dep_rule(I,O,R).
  valid_dep_path(I,O,R) :- in(I,P,_), out(O1,P,_), O != O1, 
                           dep_rule(I,O1,R1), connected(O1,I1), I != I1,
                           valid_dep_path(I1,O,R2), compose(R1,R2,R).
\end{verbatim}
\end{footnotesize}
The {\tt connected(O,I)} relation is true if the output {\tt O} shares
a data block with the input {\tt I} (implying two program blocks share
a dataflow connection from {\tt O1} to {\tt I1}):
\begin{footnotesize}
\begin{verbatim}
  connected(O,I) :- out(O,_,D), in(I,_,D).
\end{verbatim}
\end{footnotesize}
The {\tt compose(R1,R2,R)} relation implements the basic dependency
composition rules defined in the previous section:
\begin{footnotesize}
\begin{verbatim}
  compose(R1,R2,R1) :- weaker(R1,R2).
  compose(R1,R2,R2) :- weaker(R2,R1).
\end{verbatim}
\end{footnotesize}
The {\tt weaker(R1,R2)} relation encodes the ``strength'' of
dependency ordering over types (i.e., the $\preceq$ relation; see
\secref{sec:model}).  Thus, {\tt weaker(R1,R2)} is true for types {\tt
  R1} and {\tt R2} iff {\tt R1} $\preceq$ {\tt R2}.  The two {\tt
  compose} rules select the weaker relation of {\tt R1} and {\tt R2}.
If {\tt R1} is weaker than {\tt R2}, then the first {\tt compose} rule
selects {\tt R1}, and if {\tt R2} is weaker than {\tt R1}, then the
second {\tt compose} rule selects {\tt R2}. Finally, the first rule of
{\tt valid\_dep\_path} considers the case where the path is a single
program block, and the second rule considers the case where a path
consists of multiple program blocks. For the the second {\tt
  valid\_dep\_path} rule, we require {\tt O} and {\tt O1} as well as
{\tt I} and {\tt I1} to be different values, respectively, for the
case where {\tt I} and {\tt O} form a simple cycle.  Without the
inequalities, checking {\tt valid\_dep\_path} for {\tt I} and {\tt O}
would require {\tt valid\_dep\_path} for {\tt I} and {\tt O} to be
already known (from the body of the rule).
We note that workflow cycles, however, are supported by the rules.
%
The following constraint ensures that annotations are the
``strongest'' along multiple program-block paths.
\begin{footnotesize}
\begin{verbatim}
  :- dep_rule(I,O,R), valid_dep_path(I,O,R1), R != R1, weaker(R,R1).
\end{verbatim}
\end{footnotesize}
The constraint ensures there is not a stronger type between the input
{\tt I} and output {\tt O} than the one given (guessed or inferred) by
the annotation {\tt dep\_rule(I,O,R)}.

\section{Related Work}
\label{sec:related}


We focus on the PROV model, 
data provenance, 
and other workflow-based approaches: 

The PROV model \cite{prov} defines a general {\em wasInfluencedBy}
relationship with {\em wasDerivedFrom} as the main lineage
relationship between entities. PROV also defines subtypes of {\em
  wasDerivedFrom}, including {\em wasRevisionOf}, {\em
  wasQuotedFrom}, and {\em hadPrimarySource}. Although {\em DependsOn}
and {\em DerivedFrom} are similar to {\em wasInfluencedBy} and {\em
  wasDerivedFrom}, because our approach is designed for computation
via workflows, we adopt the more specific notions of dependency (i.e.,
control and data dependence) from \cite{cheney11}. Our approach is
also similar to PROV-O \cite{provonto}, which models provenance at the
schema level. We also consider compositions of dependency
annotation types, which are not considered within PROV-O.

Cui and Widom \cite{cui03} define three types of transformations for
ETL workflows---dispatchers, aggregators, and black-boxes---and for
each a set of techniques for inferring data-level lineage. 
They also define a number of specialized (i.e., a hierarchy of)
transformation types for computing data lineage. While our approach
also provides dependency types for transformations (in our case,
program blocks), the focus in \cite{cui03} is to compute data-level
workflow lineage (the input items that contributed to output items),
and does not consider the differences between dependency, derivation,
and so on. The approach used in LabelFlow \cite{alper18} is also
similar to that in \cite{cui03}, in which different types of workflow
steps are considered and used for data annotation propagation (i.e.,
arbitrary metadata attribute-value pair ``labels''). Like
\cite{cui03}, LabelFlow focuses on workflow execution by inferring
data-level labels for intermediate and final workflow data products.

Cheney \etal~\cite{cheney11} employ dependency analysis techniques
(program slicing), which are focused on calculating data dependencies
to infer ``dependency provenance'' for a query language based on the
nested relational calculus. Unlike other approaches for inferring
lineage from queries, \cite{cheney11} employs dependency analysis to
formalize the notion of lineage relationships.  Huq \etal~\cite{huq13}
describe a tool to compute data-level lineage for workflows defined as
Python scripts using Program Dependence Graphs (PDGs)
\cite{ferrante87}.  However, control dependencies are converted to
data dependencies to simplify lineage relationships for
scientists. PDGs are closely aligned with program slicing techniques,
and offer a formal interpretation of dependency also adopted by our
model.

In \cite{dey15}, data dependencies are inferred from scripts and are
then connected to YesWorkflow specifications; a prototype linking
YesWorkflow models and noWorkflow traces has been described in
\cite{pimentel2016yin}.  Our approach differs from, but complements
these approaches by explicitly supporting lineage assertions for both
control and data dependency information (among other types of
dependencies) for workflow specifications and enables validation and
inference procedures over lineage annotations.

\section{Conclusion and Future Work}
\label{sec:conclusion}

This paper defines provenance dependency types for modeling lineage
constraints within scientific workflow specifications along with a
reasoning framework that can validate dependency annotations and infer
a complete set of annotations for workflow specifications, including
the allowable choices (\emph{possible worlds}) when multiple
annotation types are possible.
We plan to extend YesWorkflow \cite{mcphillips15},
which uses annotations to declare workflow specifications for
executable scripts, with dependency annotations and the reasoning
framework described here. We also plan to develop support for
annotating subworkflows within YesWorkflow.  While the dependency
types described here cover a wide range of cases, additional types may
be needed for some workflows. For instance, although not described in
this paper, we have recently developed extensions for supporting a
{\em NotFlowsFrom} dependency type, which is needed in some
subworkflows to capture cases where subworkflow inputs are not
connected (i.e., not ``up-stream'') from subworkflow outputs. Adding
{\em NotFlowsFrom} required only minimal changes to the rules
presented in \secref{sec:prototype}. Finally, we also intend to
explore using static dependency annotations in YesWorkflow models to
infer trace-level (runtime) data lineage relationships, thus combining
our prior work in \cite{bowers12} with the reasoning framework
presented here.


\paragraph{Acknowledgements.}~\\  Work supported in part through NSF
awards OAC-1541450 and SMA-1637155.

\newpage

\bibliographystyle{alpha-initials-big}
\bibliography{main}

\end{document}